\documentclass[twoside,showpacs,reprint,superscriptaddress,amsmath,amssymb,aps,pra]{revtex4-1}
\usepackage[utf8]{inputenc}
\usepackage{amsmath}
\usepackage{amssymb}
\usepackage{graphicx}
\PassOptionsToPackage{normalem}{ulem}
\usepackage{ulem}

\makeatletter
 
 \@ifundefined{textcolor}{}
 {%
   \definecolor{BLACK}{gray}{0}
   \definecolor{WHITE}{gray}{1}
   \definecolor{RED}{rgb}{1,0,0}
   \definecolor{GREEN}{rgb}{0,1,0}
   \definecolor{BLUE}{rgb}{0,0,1}
   \definecolor{CYAN}{cmyk}{1,0,0,0}
   \definecolor{MAGENTA}{cmyk}{0,1,0,0}
   \definecolor{YELLOW}{cmyk}{0,0,1,0}
 }

\usepackage{color}
\makeatother

\begin{document}

\title{Fluctuations of the Electromagnetic Local Density of States as a
Probe for Structural Phase Switching}

\author{N. de Sousa}
\affiliation{Departamento de Física de la Materia Condensada, Universidad Autónoma
de Madrid, 28049, Madrid, Spain.}

\affiliation{Donostia International Physics Center (DIPC), Paseo Manuel Lardizabal
4, 20018 Donostia-San Sebastian, Spain.}

\author{J.J. Sáenz}

\affiliation{Donostia International Physics Center (DIPC), Paseo Manuel Lardizabal
4, 20018 Donostia-San Sebastian, Spain.}

\affiliation{IKERBASQUE, Basque Foundation for Science, 48013 Bilbao, Spain.}

\author{F. Scheffold}

\affiliation{Physics Department, University of Fribourg, Chemin du Musée 3 CH-1700
Fribourg, Switzerland.}

\author{A. García-Martín}

\affiliation{IMM - Instituto de Microelectrónica de Madrid (CNM-CSIC), Isaac Newton
8, PTM, Tres Cantos, E-28760 Madrid, Spain.}

\author{L.S. Froufe-Pérez}

\email{luis.froufe@unifr.ch}

\affiliation{Physics Department, University of Fribourg, Chemin du Musée 3 CH-1700
Fribourg, Switzerland.}
\begin{abstract}
We study the statistics of the fluorescence decay rates for single
quantum emitters embedded in a scattering medium undergoing a phase
transition. Under certain circumstances, the structural properties
of the scattering medium explore a regime in which the system dynamically
switches between two different phases. While in that regime the light
scattering properties of both phases are hardly distinguishable, we
demonstrate that the lifetime statistics of single emitters with low
diffusivity is clearly dependent on the dynamical state in which the
medium evolves. Hence, lifetime statistics provides clear signatures
of phase switching in systems where light scattering does not. 
\end{abstract}


\maketitle
The sensitivity of the spontaneous emission rate of an excited dipole
emitter to the local environment \cite{purcell1946spontaneous} makes
single-molecule spectroscopy a unique tool to sense optical and structural
properties in its surroundings on the nanoscale \cite{Moerner1999,*Weiss1999,Xie1994,Vallee2003,*Vallee2005,vallee2006fluorescence,Sauer2010}.
Control of the emission rate has been demonstrated using a variety
of well-defined structures, such as metal surfaces \cite{Chance1978},
cavities \cite{Berman1994}, photonic crystals \cite{martorell1990observation,*lodahl2004controlling},
or nanoantennas \cite{Anger2006,*Kuehn2006,*Ringler2008,*Curto2010}.
Understanding the basic physics of spontaneous emission rates in complex
media is of paramount importance for many applications (molecular
imaging techniques \cite{Moerner1999,*Weiss1999,Xie1994,Vallee2003,*Vallee2005,vallee2006fluorescence,Sauer2010},
solar cells \cite{Oregan1991}, laser technology \cite{Painter1999,gottardo2008resonance}
or single photon sources \cite{Michler2000}) which explain the increasing
interest on their statistical properties in random environments \cite{skipetrov_PRE_2006,froufe2007fluorescence,*froufe2008lifetime,Carminati_PRA_2010,Sapienza_Science_2010,*Lodahl_NJP_2011,Ruijgrok2010,Mosk_PRL_2010,sapienza2011long}.

From a fundamental point of view, the emission rate is proportional
to the number of optical modes available for emission at the position
of the emitter, i.e. proportional to the electromagnetic local density
of states (LDOS) \cite{Joulain2003definition,*Carminati_Sci_Rep_2015}.
In a complex disordered medium the LDOS presents strong fluctuations
due to dynamic conformational fluctuations of the system around the
emitter or when the emitter itself diffuses through it \cite{Xie1994,Vallee2003,*Vallee2005,vallee2006fluorescence}.
The statistical fluctuations of the LDOS \cite{skipetrov_PRE_2006,Carminati_PRA_2010,Froufe_PRA_2015}
are directly linked to the so-called $C_{0}$ speckle correlations
\cite{Shapiro_PRL_1999,Skipetrov_PRB_2000}. In absence of spatial
correlations, the averaged LDOS and the transport extinction mean
free path, $\ell$, are linked through causality Kramers-Kronig relations
\cite{Carminati2009} and the LDOS fluctuations, $C_{0}$, were predicted
to increase with the scattering strength, $\sim\ell^{-1}$ \cite{Shapiro_PRL_1999}.
However, the correlations between the emitter position and the surrounding
scatterers, due to the unavoidable excluded volume around the emitter,
makes the LDOS and its fluctuations strongly non-universal \cite{Skipetrov_PRB_2000}
and sensitive to both $\ell$ and the local correlation length\cite{Skipetrov_PRB_2000,froufe2007fluorescence,*froufe2008lifetime,Carminati_PRA_2010,Donaire_PRA_2011}.

The near-field effects on the LDOS close to a single particle are
relatively well understood \cite{Ruppin1982,*Chew1987,*carminati2006radiative,*Rolly2012,*Schmidt2012}.
In random media, when the positions of the scatterers around the emitter
are not correlated, numerical simulations show that the LDOS fluctuations
can be explained to a large extent by a single scattering statistical
model \cite{froufe2007fluorescence,*froufe2008lifetime} and are dominated
by the near-field interaction with the nearest scatterer at the scale
of the excluded volume \cite{froufe2007fluorescence,*froufe2008lifetime,Carminati_PRA_2010}.
Temporal lifetime fluctuations can then be correlated to fluctuations
in the position of the nearest scatterer and provide a suitable probe
for the dynamics of the structure around the emitter \cite{vallee2006fluorescence}.
In particular, the predicted non-Gaussian long-tailed distributions
of emission rates in disordered dielectrics \cite{froufe2007fluorescence,*froufe2008lifetime,Carminati_PRA_2010}
are compatible with experimentally measured ones \cite{sapienza2011long}.

However, similar experiments do not show such long-tailed distributions
\cite{Mosk_PRL_2010}. This result has been attributed to finite size
effects in the scatterers. Recent experiments \cite{Scalia2015} also
suggest that hydrophobic interaction between the scatterers and the
solvent in a colloidal suspension plays an important role in the description
of the decay rate and quantum yield statistics. On the other hand,
structural correlations in the disorder structure have a profound
effect in the lifetime statistical distributions \cite{Carminati_PRA_2010,deSousa_PRA_2014}.
All the reported results, show that the near-field scattering plays
an essential role in the description of lifetime statistics in disordered
media. Near field effects have also important consequences in mesoscopic
light transport \cite{Skipetrov2014,Naraghi2015}.

\begin{figure}
\includegraphics[width=0.9\columnwidth]{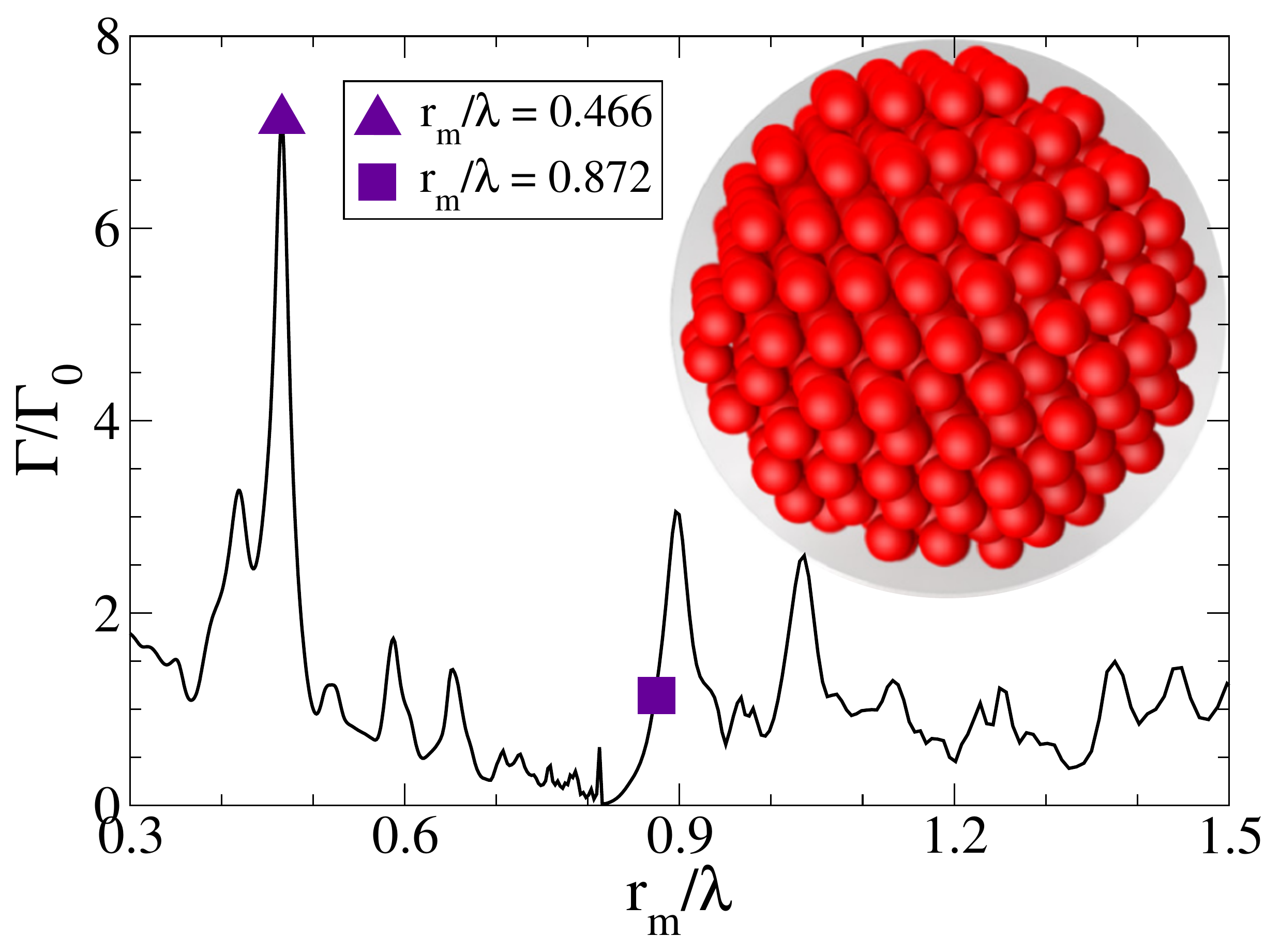}

\caption{\label{fig:fig_1}(Color online). Normalized decay rates (equal to
the LDOS normalized to the vacuum one) spectrum of a single emitter
placed at the center of a cluster at T=0 (see text for further details).
In the inset we represent the system under study. Each point scatterer
is replaced by a sphere of radius $r_{m}/2$. The translucid sphere
represents the confining sphere.}
\end{figure}

In this work we show that the statistics of emission rates in correlated
disordered media is extremely sensitive to the details of the radial
distribution function around the emitter. We analyze the emission
statistics for a single emitter embedded in a finite cluster of resonant
particles in a model system similar to that described in previous
works \cite{Carminati_PRA_2010}. However, instead of generate random
configurations of scatterers \cite{froufe2007fluorescence,*froufe2008lifetime,Carminati_PRA_2010},
we compute the emission rates as the system evolves with time under
equilibrium conditions. Assuming a standard Lennard-Jones (L-J) interaction
between particles, this system is known to present a peculiar solid-liquid-like
phase transition at finite temperature: Due to finite-size effects,
the two phases cannot coexist at the melting temperature and the whole
cluster presents an interesting dynamical behavior, switching between
an amorphous solid-like phase and liquid-like phases \cite{Briant1975,*Berry1984,*Honeycutt:1987,*Labastie1990,*Wales1994,deSousa:2016phase}.
This makes it an ideal model system to analyze the effects of local
order on the emission rates. At very low temperatures, the system
is a well ordered structure that, in the limit of infinite size, would
correspond to a Face Centered Cubic lattice. Due to this order, the
spectrum of emission rates present a strong chromatic dispersion reminiscent
of the band structure of an infinite crystal of resonant dipoles \cite{VanCoevorden1996},
including spectral windows where the emission is enhanced and pseudo-gaps
where it is dramatically inhibited \cite{Yablonovitch:1987tf,Sprik1996,*Vries1998}.
At the melting temperature, the total scattering cross section of
the system does not present significant differences between the two
phases while the emission rate jumps following the dynamic of the
system. While light scattering measurements would be blind to such
dynamical changes, the lifetime statistics would then provide a direct
signature of a phase switching behavior.

In our model system, sketched in the inset of Fig. \ref{fig:fig_1},
we consider a three-dimensional cluster of $N=515$ particles confined
inside a spherical cavity. The particles interact through a Lennard-Jones
(L-J) potential: 
\begin{equation}
V_{LJ}\left(r\right)=\varepsilon\left[\left(\frac{r_{m}}{r}\right)^{12}-2\left(\frac{r_{m}}{r}\right)^{6}\right],\label{eq:sSH_eq}
\end{equation}
where $\varepsilon$ is the depth of the potential well, $r$ is the
distance between particles and $r_{m}$ is the equilibrium distance
of the potential. The confining spherical volume is chosen in such
a way that near crystal density is achieved \cite{deSousa:2016phase}.

From the ensemble of $N$ particles, the one closest to the center
of the distribution is considered to be a point emitter. The remaining
$N-1$ particles are considered to be resonant light scatterers with
an electric polarizability, $\alpha=i6\pi/k^{3}$ (where $k=2\pi/\lambda$
is the light wavenumber and $\lambda$ the wavelength). The electrodynamic
response is obtained by using a coupled dipole method described elsewhere
\cite{froufe2007fluorescence,*froufe2008lifetime,Carminati_PRA_2010}
(which involves the solution of $3N$ self-consistent multiple scattering
equations, see appendix\ref{app:App_A}). We compute both the total scattering cross section (assuming
an external incoming plane wave) and the LDOS at the emitter position,
details of both computations are given in appendix \ref{app:App_A}. The vacuum normalized
LDOS is also the ratio $\Gamma/\Gamma_{0}$ of the emission decay
rate $\Gamma$ of a point emitter (placed at the considered position and
emitting at the considered wavelength $\lambda$) to its emission
decay rate in vacuum $\Gamma_{0}$. In Fig. \ref{fig:fig_1}, we plot
the normalized LDOS at the centre of the cluster
(after complete relaxation relazation of the structure at $T=0$), as a function of $r_{m}/\lambda$, the
ratio between the potential equilibrium distance $r_m$ and the emission wavelength $\lambda$.

The rich, peaked structure in this pseudo-spectrum (reminiscent of
the band structure of an infinite crystal of resonant dipoles \cite{VanCoevorden1996})
is a consequence of the interplay between diffraction and multiple
scattering effects of light with the crystal structure, enhanced by
the resonant character of the scatterers. We highlight two representative
points in the decay rate pseudo-spectrum: $r_{m}/\lambda=0.466$ where
the decay rate is much larger than in vacuum and $r_{m}/\lambda=0.872$,
where the decay rate is similar to the one in vacuum.

\begin{figure}
\includegraphics[width=1\columnwidth]{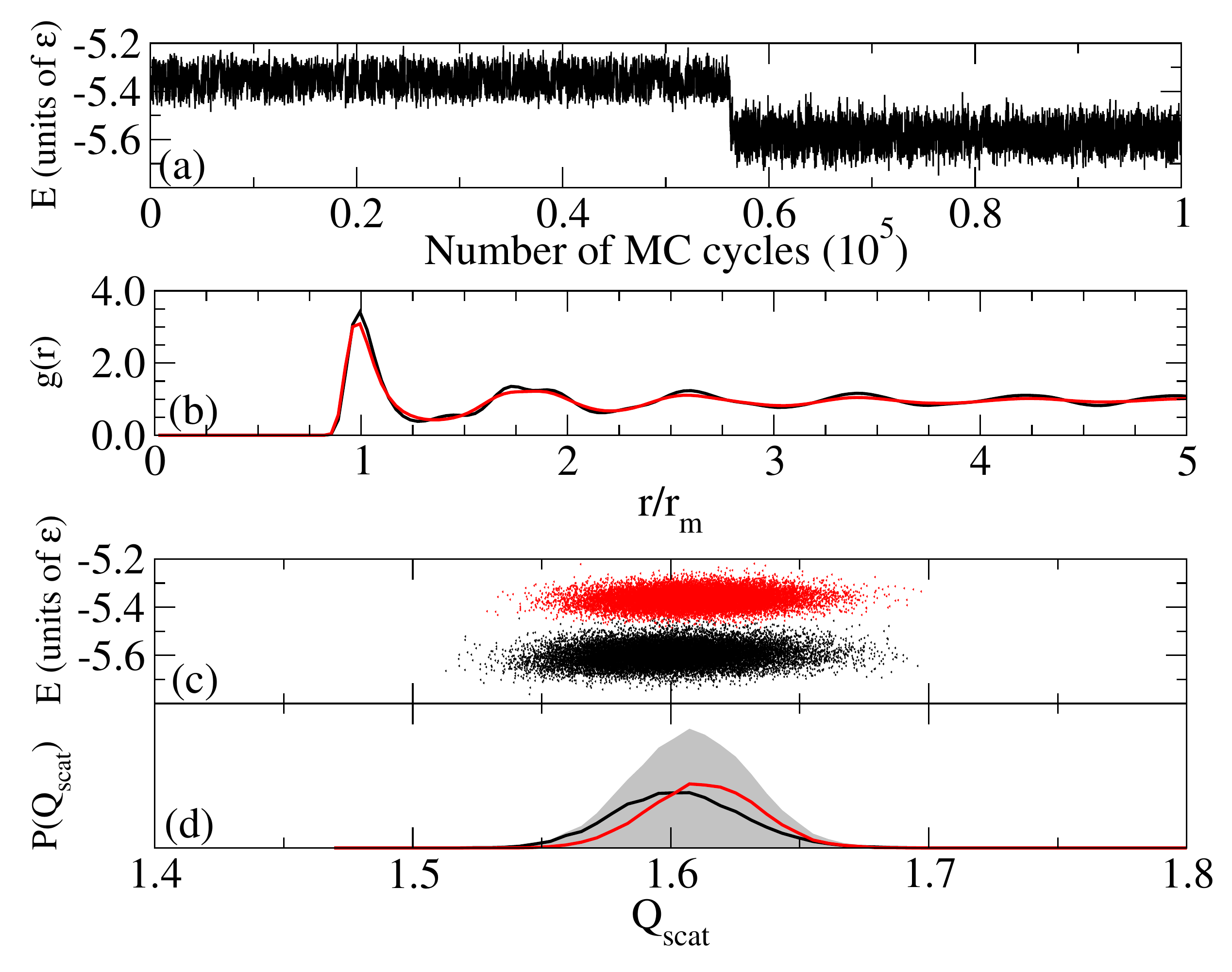}

\caption{\label{fig:fig_2}(color online). (a) Energy per particle sampling at $T=0.6$
as a function of the MC steps. (b) Corresponding pair correlation
function among scatterers $g\left(r\right)$ for the high energy branch,
in red (gray in B/W), and for the low energy branch (black). (c) Energy-scattering efficiency
sampling at the switching region ($T=0.6$, $r_{m}/\lambda=0.872$)
dots in red (gray in B/W) correspond to high energy states and black
dots to low energy states. (d) Corresponding scattering efficiencies
distributions for the upper energy branch, red (gray in B/W) curve,
and lower energy branch (black curve) corresponding to liquid and
solid phases respectively. The shaded area corresponds to the sum
of both high and low energy distributions.}
\end{figure}

In order to generate a suitable statistical ensemble at fixed temperature,
we perform standard Dynamic Monte Carlo (DMC) simulations \cite{Landau2014}
using the canonical ensemble. We depart from a crystalline structure
and perform $10^{8}$ of DMC steps (single particle moves) to thermalize the system. After
this process an extensive DMC sampling is performed computing the
scattering efficiency and LDOS for $2\times10^{4}$ configurations,
each these configurations are obtained after $10^{5}$ single-particle DMC steps.
Details of the statistical DMC simulations are given in \cite{deSousa:2016phase}.
If the temperature of the system is $\tilde T$, we define a normalized temperature 
$T\equiv K_B\tilde T/\varepsilon$, where  $K_B$ is Boltzmann's constant, and $\varepsilon$ is the
L-J potential well depth.
In particular at temperature $T=0.6\equiv T_{m}$ , the system presents the
aforementioned dynamical phase switching between low (solid-like)
and high (liquid-like) energy branches. In Fig. \ref{fig:fig_2}a
we plot the energy per particle sampling as a function of the number of DMC cycles
and a switch event from high to low energy is clearly observed. The
average of the self-diffusion coefficients was found to largely vary
from the liquid-like to the solid-like phases, providing an unambiguous
signature of the actual phase state \cite{deSousa:2016phase}. Interestingly,
the same simulations showed that the pair correlation function $g\left(r\right)$
is essentially the same for both phases, as shown in Fig.
\ref{fig:fig_2}b \cite{deSousa:2016phase}. This indicates that the
system switches from liquid to an amorphous solid phase rather than
crystal-like and suggest that light scattering experiments could not
be sensitive to this subtle dynamical switching. As a matter of fact,
this is consistent with our numerical results shown in Fig. \ref{fig:fig_2}c
where we present the energy sampling versus the computed normalized
scattering cross-section, $Q_{scat}$ (scattering efficiency) for
$r_{m}/\lambda=0.872$. To guide the eye, points corresponding to
high and low internal energy are rendered in different colors. Integrating
the sampling in energy, we obtain scattering efficiency histograms
as shown in Fig. \ref{fig:fig_2}d. Differences in $Q_{scat}$ histograms
corresponding to high and low energy phases can be hardly distinguished.
The $Q_{sscat}$ histogram obtained by considering all the values of $Q_{scat}$
for all possible energies (shaded gray area in Fig.\ref{fig:fig_2}d)
shows a single peak and no signature of the two-state switching.

It is well known that positional correlations between scatterers can
strongly affect the wave transport properties, i.e. the transport
mean free path, in bulk disordered media. They are responsible, for
example, of the large conductivity of liquid metals \cite{ashcroft1966structure},
the cornea transparency \cite{cox1970transparency}, the strong chromatic
dispersion in colloidal suspensions \cite{rojas2004photonic,*Scheffold_colloids_2007}
and amorphous photonic materials \cite{Reufer2007,*Garcia2007photonic,*Muller_Scheffold_Adv_Opt_Mat_2014}
or natural structural coloration \cite{Liew2011,*Cortese2015}. The
correlations in wave transport through a translational invariant system
are encoded in the pair correlation function $g\left(r\right)$. As
expected, we conclude that light scattering experiments would not
provide a way to distinguish between phases in the switching regime
due to the indistinguishability of the $g\left(r\right)$ in the different
dynamical regimes.

\begin{figure}
\includegraphics[width=1\columnwidth]{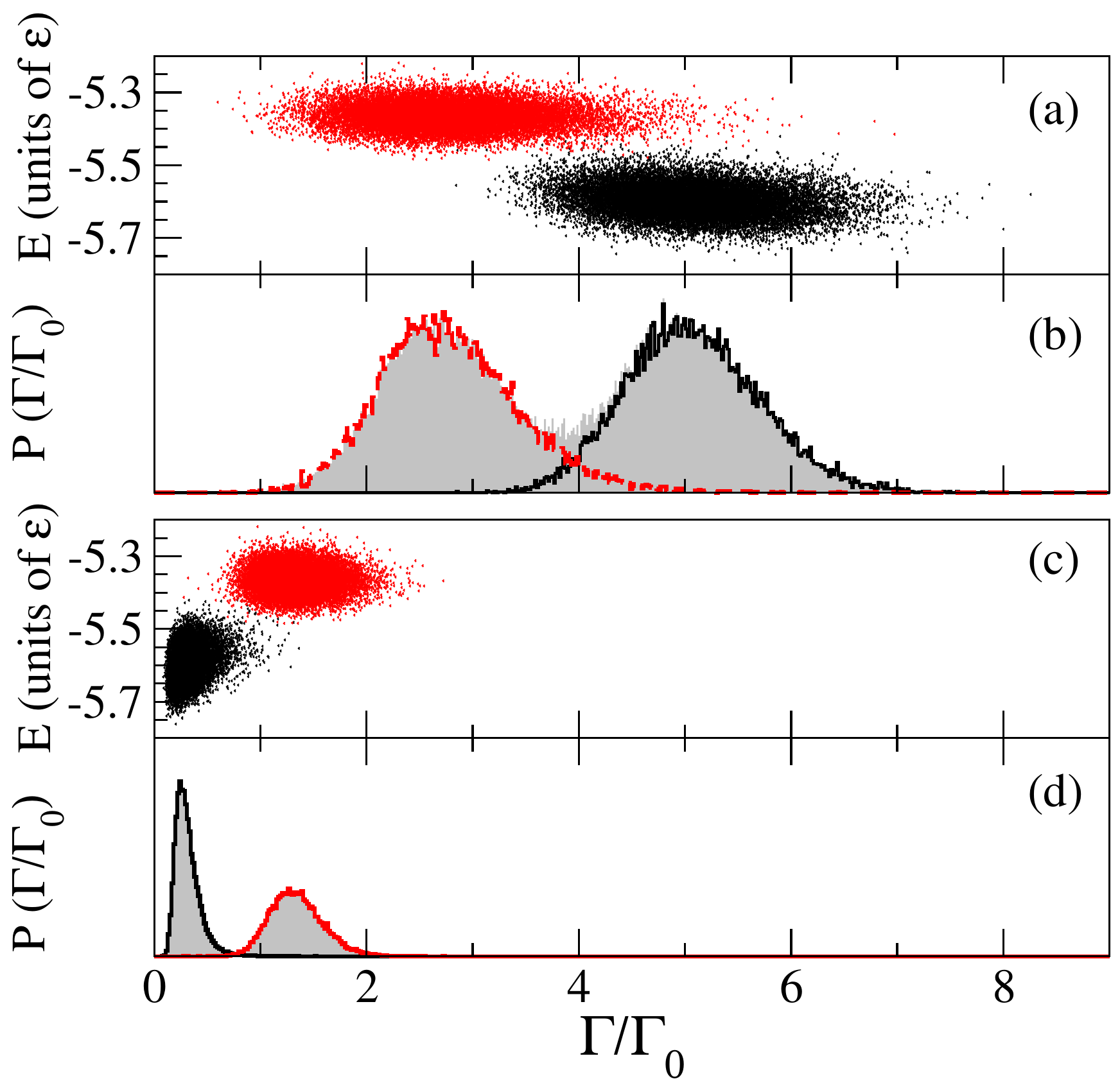}

\caption{\label{fig:fig_3}(color online). Energy per particle - decay rate sampling at the
switching region ($T=0.6$) for different ratios $r_{m}/\lambda$,
(a) $r_{m}/\lambda=0.466$; (c) $r_{m}/\lambda=0.872$. In the panels
(b,d), the corresponding decay rates distributions are shown after
integrating in energies: black histograms correspond to the solid
phase, red (gray in B/W) ones to liquid phase, and the shaded gray
areas to total measured decay rates (sum of both solid and liquid
distributions). }
\end{figure}

However, emission decay rate statistics (or LDOS) shows clear signatures
of the phase-switching regime. In Fig. \ref{fig:fig_3}, panels (a,c),
we present an energy-decay rate sampling performed at $T=0.6$ at
two different ratios of the interaction potential characteristic length
to emission wavelength $r_{m}/\lambda$ (the ones highlighted in Fig.
\ref{fig:fig_1}). In the present model, the point emitter is chosen
to be located at the position of the interacting particle closest
to the origin. In this way, the dynamics of the emitter and the remaining
scatterers is indistinguishable. The direction of the radiating dipole,
is random and considered to be uniformly distributed among the whole
$4\pi$ angles. We have verified that, despite the fact that the spatially
averaged self-diffusion constant varies by a factor three between
the two phases, the particle located initially at the center of the
cluster hardly diffuses along the DMC calculation. Hence, we can consider
the emitter as a low diffusivity one.

 On the other hand, as discussed in more detail in Appendix \ref{app:App_A}, we calculate the emission decay rates
of the dipole emitter considering all the multiple scattering in the system. We nevertheless do not take into account any 
far-field radiation delay due to radiation trapping. Those effects might be present and would be caused by coupling to
long-lived modes into the sample. However, it can be argued that the sample is in the diffusive regime (see Appendix \ref{app:App_B})  where such long-lived modes 
should be rare. In fact, as demonstrated in \cite{Skipetrov2014}, at least in collections of uncorrelated disordered point scatterers,
such long-lived modes do not exist.

The emission rates evolve with time following the fast structural
changes in the dynamic coexistence region. As it can be observed in
panels (a,c) of Fig. \ref{fig:fig_3}, the decay rate distributions
corresponding to lower energetic levels (solid phase, black dots/lines)
are different from the higher energetic levels (liquid phase, red
dots/lines). In particular its average values and fluctuations
are appreciably different. Collecting all the emission rates in a
histogram results in the statistical distributions of emission rates
shown as shaded gray aread in Fig. \ref{fig:fig_3}(b,d).
For the selected working wavelengths, the distributions are always
bimodal, showing a clear signature of the phase switching.

\begin{figure}
\includegraphics[width=1\columnwidth]{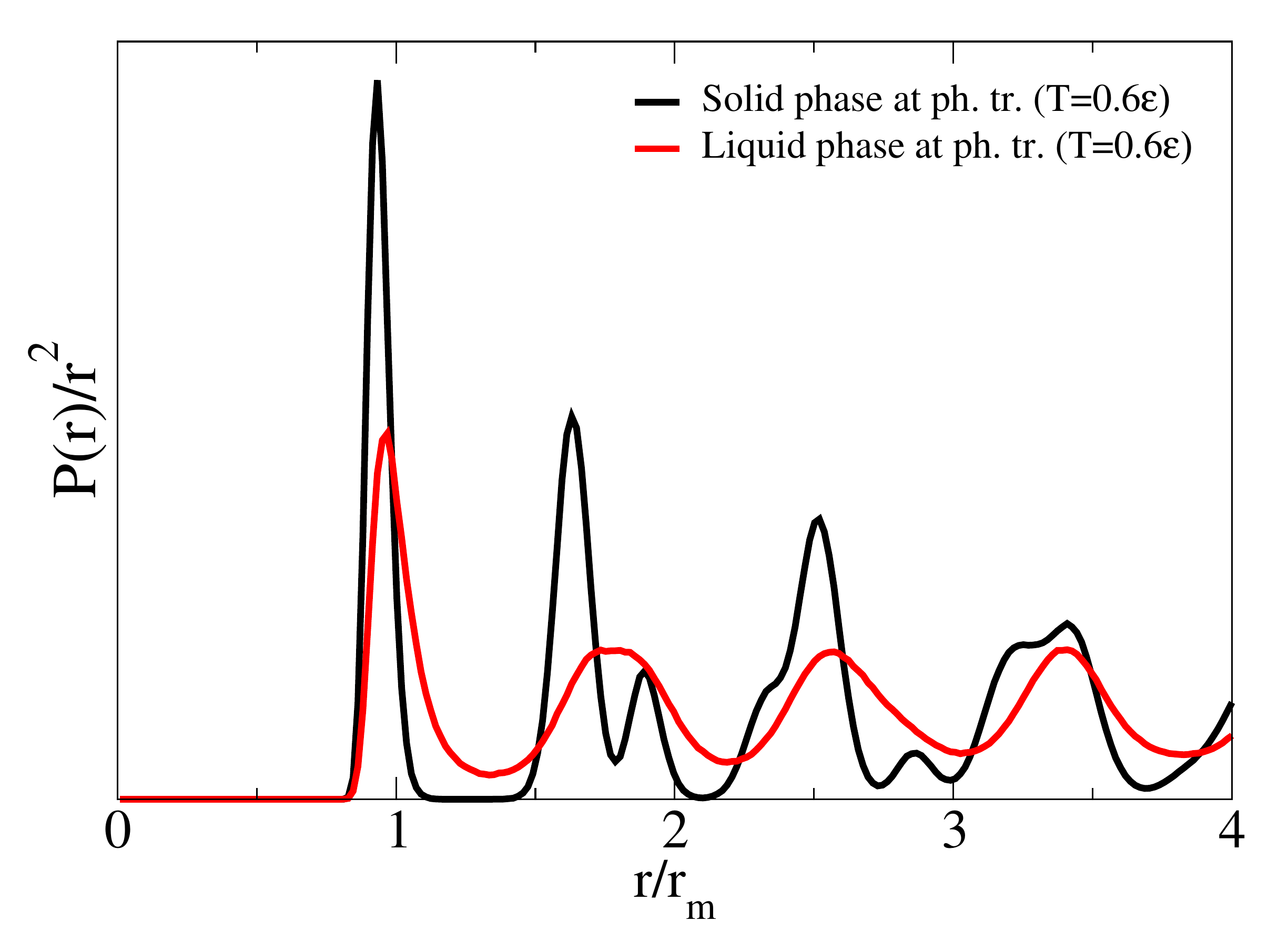} \caption{\label{fig:fig_4}(color online). Emitter radial distribution function (ERDF)
for scatterers surrounding the emitter at phase switching region ($T=0.6$).
Black and red (gray in B/W) curves represents the solid and liquid
region respectively.}
\end{figure}

In order to clarify the origin of the statistical signatures of phase
switching in single emitter decay rates, we analyze in the following
the normalized {\em{emitter radial distribution function}} (ERDF) of scatterers around
the emitter. The ERDF is defined as the probability of finding a particle
at a distance $r$ from the emitter $P\left(r\right)$ normalized
to the probability in absence of any correlation ($\propto r^{2}$).
In this paper we make a distinction between ERDF and the pair correlation
function. While for $g\left(r\right)$ we consider all pairs of scatterers,
we reserve the term ERDF only for the distributions of distances between
the emitter and the scatterers. In a translationally invariant system,
both distributions should be the same since we consider the emitter
to have the same dynamics as the scatterers. Nevertheless, as shown in
the next paragraphs, in our relatively small and strongly confined
system, the emitter, despite being subjected the same interaction
potential, behaves in a singular way as compared to the remaining
scatterers pairs because it is placed very close to the center and,
at the temperatures of interest, remains close to its initial position
during the course of the simulation. 

In contrast to the pair correlation function $g\left(r\right)$, the
ERDF at constant temperature shows dramatic variations that follow
the phase switching. In Fig. \ref{fig:fig_4}, we show the ERDF at
$T=0.6$ calculated in the low energy regime, or amorphous solid phase,
and in the high energy regime or liquid phase. As can be observed
in this figure, the solid phase exhibits a richer peak structure than
the liquid phase. This fact might be related to a better layering
of the structured around its center. In particular, we observe that
the probability of finding particles close to the emitter, represented
by the height of the first peak, is much higher in the solid phase
than in the liquid one.

With the above considerations, the physical picture we devise is as
follows. The ensemble averaged scattering cross section
is determined by the pair correlation function $g\left(r\right)$.
Hence, for identical scatterers, similar $g\left(r\right)$ shall
lead to similar scattering properties. The dynamics of light emission
by single emitters, however, is controlled not only by the multiple
scattering properties of the whole ensemble but also by the distribution
of scatterers around the emitter, in turn described by the ERDF 
\cite{froufe2007fluorescence,*froufe2008lifetime}. Hence,
if we have a system showing disparate ERDFs for a slowly diffusing
emitter, the lifetime emission statistics of such an emitter can be
controlled by the ERDF variations even when the $g\left(r\right)$
remains almost unchanged.

In summary, we have presented a model system of interacting light
scatterers that present a solid-liquid phase transition. In the case
where the system is relatively small (few hundreds of scatterers)
and strongly confined, the system presents a phase switching regime
where it switches between phases in its entirety for a certain range
of temperatures. We have shown that, due to the fact that $g(r)$
functions are nearly indistinguishable between both phases, static
light scattering experiments would not be able to discriminate between
phases in the switching regime.

Strikingly, we find that single emitter decay-rate statistics shows
strong signatures of the phase switching regime. We have correlated
this behavior to the difference in the radial distribution functions between scatterers and
the emitter position which, in turn, might also be attributed to differences
in the self-diffusion of scatterers between both phases. Therefore,
this could be experimentally verified in an experiment performed using
emitters with low diffusivity.

The system we have considered in this work presents an illustration
of one deep difference between light scattering and light emission.
Apart from the fundamental implications of this effect, it might be
used as a tool for monitoring subtle thermodynamical behaviors in
complex systems with sizes comparable with the wavelength of the light
source employed in the experiment.

\appendix

\section{LDOS and Total Scattering Cross Section\label{app:App_A}}

In this appendix we present the expressions used for evaluate the
electromagnetic local density of states (LDOS) and the total scattering
cross sections.

Here we consider a particular frequency ($\omega=\omega_{0}$) and
an associated particular wave number ($k=k_{0}=\omega_{0}/c$) at
which dipoles are in resonance with the electromagnetic radiation,
meaning that the polarizability is now given by $\alpha=i6\pi/k_{0}^{3}$.

The electric field at some position $\mathbf{r}$, generated by the
presence of a dipole emitter $\boldsymbol{\mu}$ at some position
$\mathbf{r}'$ can be obtained by operating the Green tensor over
the dipole. This is expressed as:

\begin{equation}
\mathbf{E}\left(\mathbf{r}\right)=\frac{k^{2}}{\epsilon_{0}}\mathbf{G}_{0}\left(\mathbf{r},\mathbf{r}'\right)\cdot\boldsymbol{\mu},
\label{field_dipole}
\end{equation}

$\epsilon_{0}$ being the permittivity of vacuum.

The Green tensor is given by \cite{novotny2012principles}: 
\begin{equation}
\begin{split}\mathbf{G}_{0}\left(\mathbf{r},\mathbf{r}'\right)= & \frac{e^{ikR}}{4\pi R}\left[\left(1+\frac{ikR-1}{k^{2}R^{2}}\right)\mathbb{I}+\right.\\
 & \left.+\left(\frac{3-3ikR-k^{2}R^{2}}{k^{2}R^{2}}\right)\hat{\mathbf{R}}\otimes\hat{\mathbf{R}}\right],
\end{split}
\end{equation}

where $R$ is the modulus of the vector $\mathbf{R}=\mathbf{r}-\mathbf{r}'$,
$\hat{\mathbf{R}}\otimes\hat{\mathbf{R}}$ denotes the outer product
of $\hat{\mathbf{R}}=\mathbf{R}/R$ by itself and $\mathbb{I}$ is
the unit dyadic.

For a system formed by $N$ dipole scatterers,
the total electric field at some position $\mathbf{r}$ (outside any scatterer) is given by:

\begin{equation}
{\mathbf{E}\left(\mathbf{r}\right)=\mathbf{E}_{ext}\left(\mathbf{r}\right)+\frac{k^{2}}{\epsilon_{0}}\sum_{{n=1}}^{N}{\mathbf{G}_{0}\left(\mathbf{r},\mathbf{r}_{n}\right)}\mathbf{p}_{n},}\label{eq:electric_field}
\end{equation}

where $\mathbf{E}_{ext}\left(\mathbf{r}\right)$ is the external electric field at the considered position,
$\mathbf{r}_{n}$ is the position of the $n$-th scatterer, and $\mathbf{p}_{n}$ is the induced dipole located at $\mathbf{r_{n}}$.

Induced dipoles, $\mathbf{p}_{n}=\epsilon_{0}\alpha\mathbf{E}_n$, are obtained by self-consistently solving the set of $3N$ equations 
relating the total incoming field exciting the $n$-th dipole $\mathbf{E}_{n}$
with the external field and the field radiated from
the remaining induced dipoles, that is proportional to the total incoming fields inpinging onto each of the remaining induced dipoles:
\begin{equation}
\mathbf{E}_{n}=\mathbf{E}_{ext}\left(\mathbf{r}_n\right)+k^2\alpha\sum_{m\ne n} \mathbf{G}_{0}\left(\mathbf{r}_n,\mathbf{r}_{m}\right)\mathbf{E}_{m}.
\label{DDA_eqs}
\end{equation}
Equation(\ref{DDA_eqs}), describes the coupled dipole method \cite{purcell1973scattering}.

The second term on the right hand side of eq.(\ref{eq:electric_field}) is the scattered field

\begin{equation}
\mathbf{E}_{s}\left(\mathbf{r}\right)=k^{2}\alpha\sum_{{n=1}}^{N}{\mathbf{G}_{0}\left(\mathbf{r},\mathbf{r}_{n}\right)}\mathbf{E}(\mathbf{r_{n}}).
\label{eq:scattering_field}
\end{equation}

If the external field is given by eq.(\ref{field_dipole}), the total field scattered by the collection of 
scatterers can then be calculated afted solving eq.(\ref{DDA_eqs}) with this external field. The normalized spontaneous decay rate $\Gamma/\Gamma_{0}$ of a dipole emitter
$\boldsymbol{\mu}$, in the weak coupling regime, is given by \cite{novotny2012principles}:

\begin{equation}
\frac{\Gamma}{\Gamma_{0}}=1+\frac{6\pi\epsilon_{0}}{\left|\boldsymbol{\mu}\right|^{2}k^{3}}\Im[\boldsymbol{\mu}^{*}\cdot\mathbf{E}_{s}\left(\mathbf{r}'\right)],\label{decay_rate}
\end{equation}
where $\Im$ stands for imaginary part, and $\Gamma_{0}$ is the emitter's free space decay rate. 

 In order to compute the scattering cross section $\sigma_{scat}$, we consider an incoming plane wave as the external field
$\mathbf{E}_{ext}\left(\mathbf{r}\right)=\mathbf{E}_0 \exp \left( \mathbf{k}\cdot\mathbf{r}\right)$. After solving eq.(\ref{DDA_eqs}) with this external field, 
the induced dipoles are obtained and the total scattering cross seciton of the system can be written in terms of the induced dipoles $\mathbf{p}_n$ as
\cite{deSousa2016magneto}:

\begin{equation}
\sigma_{scat}= \frac{k^{3}}{\epsilon_{0}^{2}\left|\mathbf{E}_{0}\right|^{2}}\sum_{n,m=1}^{N}
\mathbf{p}_{n}^{*}\cdot\Im \left[ \mathbf{G}_0\left(\mathbf{r}_{n},\mathbf{r}_{m}\right) \right]\mathbf{p}_{m}.
\end{equation}

\section{Transport regime\label{app:App_B}}

We used a set or resonant electric point dipoles throughout the manuscript.
An important question that might arise is whether the system is in
the quasi-ballistic, diffusive or localization regimes.

Considering the standard diffusion theory, the transport mean free
path $\ell_{tr}$, in the absence of absorption and anisotropic scattering,
can be taken as 
\begin{equation}
\ell_{tr}^{-1}=\rho\sigma_{scat}^{(p)}\label{eq:B10}
\end{equation}
Where $\rho$ is the density of scatterers and $\sigma_{scat}^{(p)}$ is the single scatterer scattering cross section. The density $\rho$ 
has been taken to be \cite{deSousa:2016phase} $\rho=1.07r_{m}^{-3}$. On the other hand, the scattering cross section at resonance is given by
\begin{equation}
\sigma_{scat}^{p}=6\pi k^{-2}\label{eq:B20}
\end{equation}

We will now estimate both the optical thickness $b\simeq R/\ell_{tr}$ and
$k\ell_{tr}$ for the cluster of radius $R$ formed by $N=515$ scatterers.
Considering that 
\begin{equation}
R=\left(\frac{3N}{4\pi\rho}\right)^{1/3}\textrm{,}\label{eq:B30}
\end{equation}
and combining eq. (\ref{eq:B10}-\ref{eq:B30}), we obtain an optical thickness

\begin{align}
b\simeq R/\ell_{tr} & =\left(\frac{3N}{4\pi}\right)^{1/3}\frac{3}{2\pi}\rho^{2/3}\lambda^{2}\simeq2.48\left(\frac{\lambda}{r_{m}}\right)^{2}\nonumber \\
 & \simeq\left\{ \begin{array}{l}
3.27\qquad\textrm{ for }r_{m}/\lambda=0.872\\
11.44\qquad\textrm{for }r_{m}/\lambda=0.466
\end{array}\right.\label{eq:B40}
\end{align}
Also, we get 
\begin{equation}
k\ell_{tr}=\frac{4\pi^{2}}{3}\left(\frac{r_{m}}{\lambda}\right)^{3}\simeq\left\{ \begin{array}{l}
27.81\qquad\textrm{for }r_{m}/\lambda=0.872\\
4.20\qquad\textrm{ for }r_{m}/\lambda=0.466
\end{array}\right.\label{eq:B50}
\end{equation}

We can conclude from the above considerations that the system can
not localize due to the large values of $k\ell_{tr}$, and that it
is well, though not very deep, in the diffusive regime due to its
relatively large optical thickness.

Of course, the ratio of the transport time to the natural decay rate
of the emitter $\tau_{0}=\Gamma_{0}^{-1}$ will depend on the chosen
emitter. In \cite{sapienza2011long}, it was argued that in a highly
scattering system with $k\ell_{tr}\simeq9.4$, the transport time ($\sim ps$)
was much smaller than the fluorescence typical time of organic dyes
($\sim ns$). We conclude hence that, despite the strong scattering
in the proposed samples, experiments using state of the art techniques
should be feasible. 
\begin{acknowledgments}
This research was supported by the Spanish Ministry of Economy and
Competitiveness through grants FIS2012-36113-C03, FIS2015-69295-C3-3-P,
and MAT2014-58860-P, and by the Comunidad de Madrid (Contract No.
S2013/MIT-2740). F.S. and L.S.F.-P. acknowledge funding from the Swiss
National Science Foundation through the National Centre of Competence
in Research Bio-Inspired Materials. 
\end{acknowledgments}

\bibliography{biblio}

\end{document}